# Lateral transport properties of thermally excited magnons in yttrium iron garnet films

X. J. Zhou,[1] G. Y. Shi,[1] J. H. Han,[1] Q. H. Yang,[2] Y. H. Rao,[2] H. W. Zhang,[2] L. L. Lang,[3] S. M. Zhou,[3] F. Pan,[1] and C. Song[1,a]

[1]Key Laboratory of Advanced Materials (MOE), School of Materials Science and Engineering, Tsinghua University, Beijing 100084, China.

[2]State Key Laboratory of Electronic Films and Integrated Devices, University of Electronic Science and Technology of China, Chengdu 610054, China.

[3]School of Physics Science and Engineering, Tongji University, Shanghai 200092, China.

Spin information carried by magnons is attractive for computing technology and the development of magnon-based computing circuits is of great interest. However, magnon transport in insulators has been challenging, different from the clear physical picture for spin transport in conductors. Here we investigate the lateral transport properties of thermally excited magnons in yttrium iron garnet (YIG), a model magnetic insulator. Polarity reversals of detected spins in non-local geometry devices have been experimentally observed and are strongly dependent on temperature, YIG film thickness, and injector-detector separation distance. A competing two-channel transport model for thermally excited magnons is proposed, which is qualitatively consistent with the spin signal behavior. In addition to the fundamental significance for thermal magnon transport, our work furthers the development of magnonics by creating an easily accessible magnon source with controllable transport.

---

[a]E-mail: songcheng@mail.tsinghua.edu.cn

Magnons (or spin waves), are the collective quasiparticle excitations of magnetic order in magnetic materials.[1,2] And magnonics, a flourishing field in spintronics which focuses on magnons, has garnered wide interest both as fundamental science and in technical applications.[3,4] In particular, innovative research is ongoing in magnon-based computing technology, representing by magnons transport and data processing.[5–9] Consequently, convenient sources of magnons with controllable transport properties are greatly needed to promote development of magnon-based computing circuits. In addition to conventional spin pumping via microwave magnetic fields,[10–12] magnons in magnetic insulators (MIs) can be excited both electrically[13–18] and thermally.[13,19,20] The spin Hall effect is involved both in magnon generation and detection in electrically excited materials.[13–18] Moreover, practical progress in devices has utilized electrically excited magnons,[15] benefiting from previous transport property research in this topic.[13,14]

Regarding thermally excited magnons (TEMs), a temperature gradient along the longitudinal direction gives rise to magnon accumulation at the opposite side of the magnetic channel, for example between a surface and a magnetic film/substrate interface.[21] Considering the distinct source of TEMs compared with their electrical counterpart, there is an attractive aspect in pursuing thermal excitation. More importantly, as TEMs have been extensively cited in the spin Seebeck effect,[22–25] the exploration of TEM transport in MIs both improves magnon control and advances the fields of magnonics and spin caloritronics.[26] In this work, we investigate the lateral transport properties of TEMs in yttrium iron garnet (YIG) films, which are model MIs with uniquely low magnetic damping.[27] The experimental results presented show that once thermally excited in the injector, the spin sign of accumulated thermal magnons in the detector can change in the non-local geometry. It is found to be highly dependent on temperature ($T$), YIG film thickness ($t$), and injector-detector separation distance ($d$). And a possible model is proposed to understand the distinct transport of TEMs in YIG.

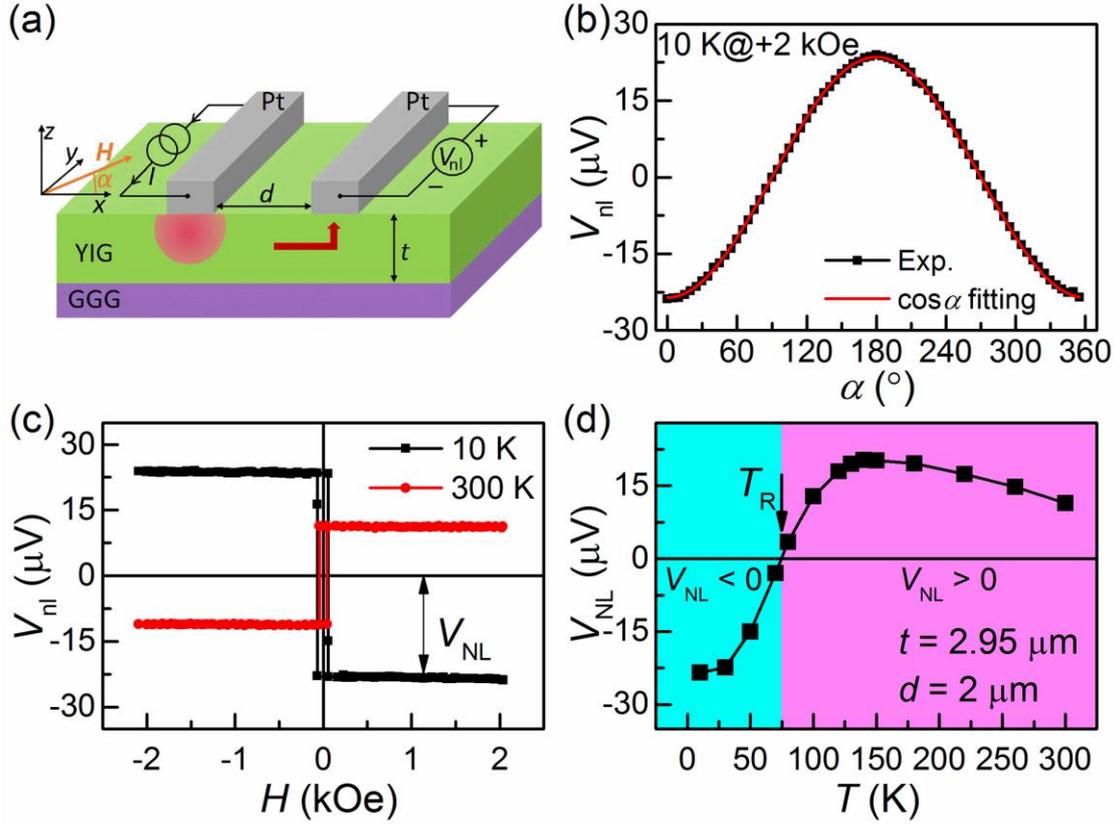

FIG. 1. (a) Sketch of the experimental configuration used to investigate TEM transport. (b) The angular dependence of $V_{nl}$ measured at 10 K. The red line is the fit to a cos$\alpha$ function. (c) $V_{nl}$ measured at 10 K (black) and 300 K (red) versus sweeping field ($H//x$). (d) Temperature dependence of $V_{NL}$ with the polarity reversal temperature labeled as $T_R$.

Bi-layer YIG/Pt samples were fabricated. YIG films with thicknesses 4 μm, 2.95 μm, and 440 nm were grown on gadolinium gallium garnet (GGG) substrates by liquid phase epitaxy, and 80 nm thick YIG films were prepared by pulsed laser deposition. A 10 nm Pt layer, fixed for all devices, was deposited by magnetron sputtering. Isolated Pt strips serving as injectors and detectors as illustrated in Fig. 1(a), defined by e-beam lithography and ion beam milling, were 500 μm long and 1 μm wide. Measurements were carried out in a cryostat at temperatures between 10 and 300 K, in which a magnetic field could be applied. Thermal magnons were excited beneath the injector at a constant current of 2 mA, and the non-local voltage was measured in the detector. More specifically, when a current (*I*) is introduced in the left Pt strip as shown in Fig. 1(a), the Joule heat generated in the injector produces temperature gradients $\nabla T$ along both the *z* and *x* axes. A TEM current is created in the surrounding YIG from the longitudinal spin Seebeck effect (LSSE) in YIG/Pt.[23] The magnons diffuse to the right Pt strip and the spin current is absorbed and induces a voltage $V_{nl}$ along the *y* axis via the inverse spin Hall effect (ISHE). Additionally, an external magnetic field (***H***) is applied parallel to the *x* axis to control the magnetization of the YIG, and subsequently modulate TEM detection.[13]

The first device tested had a YIG thickness of 2.95 μm and a separation distance of 2 μm. As displayed in Fig. 1(b), the in-plane angle-dependent $V_{nl}$ at 10 K is well fitted by a cos*α* relationship at *H* = 2 kOe. It verifies that the magnons detected are excited thermally rather than electrically, which would follow a distinct cos$^2$*α* fit.[13] In the meantime, the angular dependence keeps constant once YIG is saturated (*H* > 100 Oe) and quadratic relationship between $V_{nl}$ and *I* confirms the thermal origin as $\nabla T$ is proportional to $I^2$ (see Fig. S1 in the supplementary material). In particular, a measurable contribution from electrically excited magnons is observed in similar structures.[13,14] Enhanced scattering from the inevitable fabrication induced surface damage and interface disorder is regarded as a potential constraint on electrical excitation. Thus, we can safely attribute the spin information detected to a single TEM source.

Considering the orthogonal force relation in the ISHE, the magnetic field is always applied collinearly to the $x$ axis to obtain a maximum $V_{nl}$ in the detector. Results at low temperature (10 K) and high (300 K) are shown in Fig. 1(c). The sharp flip at small field and quick saturation of $V_{nl}$ correspond to the YIG magnetization reversal with sweeping $H$. The magnitude of $V_{nl}$ is defined as $V_{NL}$, the saturated value of $V_{nl}$ at positive magnetic field. And $V_{NL}$ at 10 K is also found proportional to $I^2$ (Fig. S1, supplementary material), reaffirming the TEM finding. Unexpectedly, $V_{NL}$ changes its sign from negative at 10 K to positive at 300 K. To get a full survey of the phenomenon, the temperature dependence is measured, summarized in Fig. 1(d). A clear $V_{NL}$ sign reversal is observed near 75 K, labeled as $T_R$, below which $V_{NL} < 0$ while $V_{NL} > 0$ above this to 300 K. It seems to contradict an LSSE study in YIG/Pt, where the SSE voltage polarity keeps unchanged from 10 to 300 K.[28] Our measurements of local thermal voltage ($V_L$) caused by LSSE in the injector remains positive despite the sign reversal in the detector (see Fig. S2, supplementary material), agreeing with conventional LSSE theory. Thus, the spin orientation of the TEMs in the detector could not be simply ascribed to proximate heating, which would only affect the magnitude. More importantly, the appearance of sign reversal in $V_{NL}$ strongly implies potential competition in TEM lateral transport.

Devices having a larger $t$ of 4 μm and distinctly smaller $t$ values of 440 and 80 nm were also investigated with the same separation distance of 2 μm. As shown in Fig. 2, the 4 μm device data are similar to the 2.95 μm device, where a clear sign reversal is seen at a somewhat lower $T_R$. However, the situation is dramatically different in small $t$ devices. With $t$ values of 440 and 80 nm, $V_{NL}$ remains negative over the entire temperature range. It is evident that the spin sign is highly dependent on the YIG thickness. Considering that $V_{NL}$ is determined by the magnon spin accumulation in the detector, TEM transport plays a critical role. To obtain further understanding, we investigate $V_{NL}$ dependence on separation distance $d$.

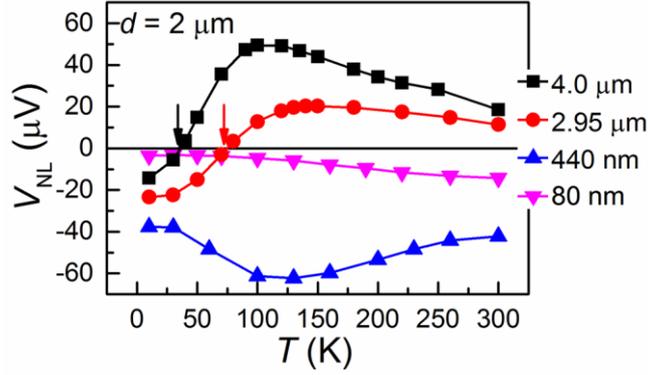

FIG. 2. Temperature dependence of $V_{NL}$ at four YIG thicknesses. The separation distance is fixed at 2 μm.

Devices with $t$ of 4 μm and 2.95 μm can be regarded as thick YIG films relative to $d$ and thick film devices exhibit the polarity reversal. Figures 3(a) and 3(b) show $V_{NL}$ versus $T$ data at different $d$. All devices with $t$ of 4 μm [Fig. 3(a)] show $V_{NL}$ polarity reversals from positive to negative as temperature decreases from 300 to 10 K. Moreover, the temperature interval where $V_{NL}$ is negative extends gradually with increasing $d$. This effect is more prominent in the thinner 2.95 μm YIG devices as presented in Fig. 3(b). For $d$ spacings of 2 and 3 μm, $T_R$ is elevated by 40 and 85 K compared with 4 μm thick devices with identical $d$. More importantly, at 2.95 μm thickness and a separation distance of 4 μm, $V_{NL}$ remains negative across the whole temperature range. So far, lateral transport of TEM in YIG is more clear that lower temperature, thinner YIG films, and larger separation distances tend to promote negative $V_{NL}$. Opposite spin accumulation or positive $V_{NL}$ is more easily detected when those conditions are reversed.

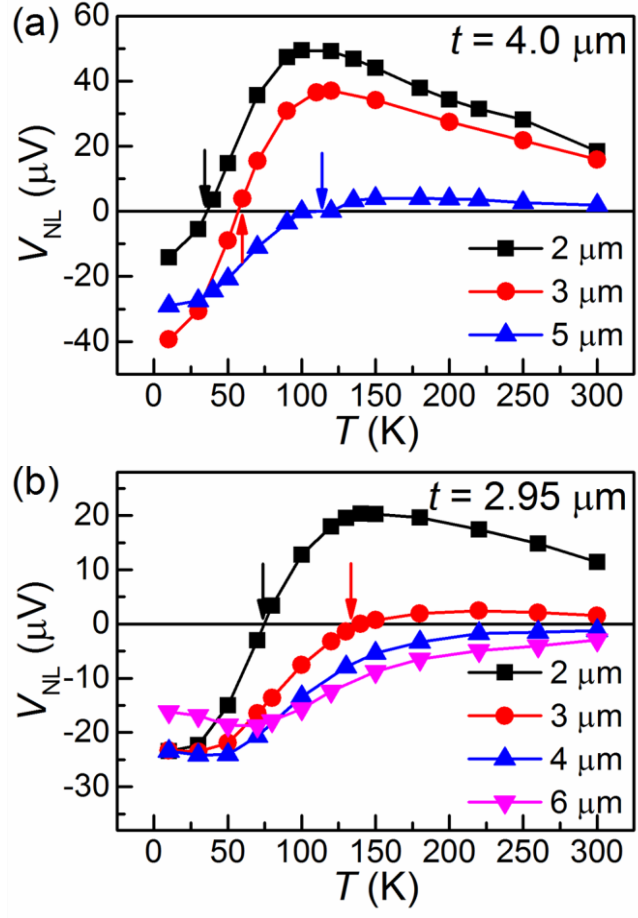

FIG. 3. Temperature dependence of $V_{NL}$ at various separation distances in devices with YIG thickness of 4 μm (a) and 2.95 μm (b).

When $t$ is significantly smaller than $d$, such as in the devices with 440 and 80 nm YIG thicknesses [Fig. 2], a dominant negative $V_{NL}$ signal is seen at all temperatures. A number of 80 nm thick devices were fabricated with different $d$ spacings. Data on $V_{NL}$ versus $T$ for these is shown in Fig. 4(a). As expected, $V_{NL}$ remains negative at all temperatures and its magnitude gradually diminishes as $d$ increases. Displaying this as a series of $V_{NL}$ versus $d$ curves at different temperatures in Fig. 4(b), a set of single exponential decay functions $V_{NL} = V_{NL}^{S} e^{-d/\lambda}$ fit the data,[19] where $\lambda$ is the magnon relaxation length and $V_{NL}^{S}$ are distance-independent prefactors. The excellent match of the $V_{NL}$ fitting to the data in Fig. 4(b) indicates that the spin signal is dominated by magnon relaxation rather than diffusive transport, which would follow a $1/d$ behavior.[13] The different temperature-dependent values of $\lambda$ are shown in the Fig. 4(b) inset. The $\lambda$ magnitudes obtained from 80 nm devices are comparable to the ~10 μm reported by Cornelissen *et al.*[13,29] and the spin diffusion length in *n*-GaAs,[30] but are obviously smaller than the ~50 μm found by Giles *et al.*.[19] This difference may be related to the YIG thickness and it is beyond the scope of our discussion.

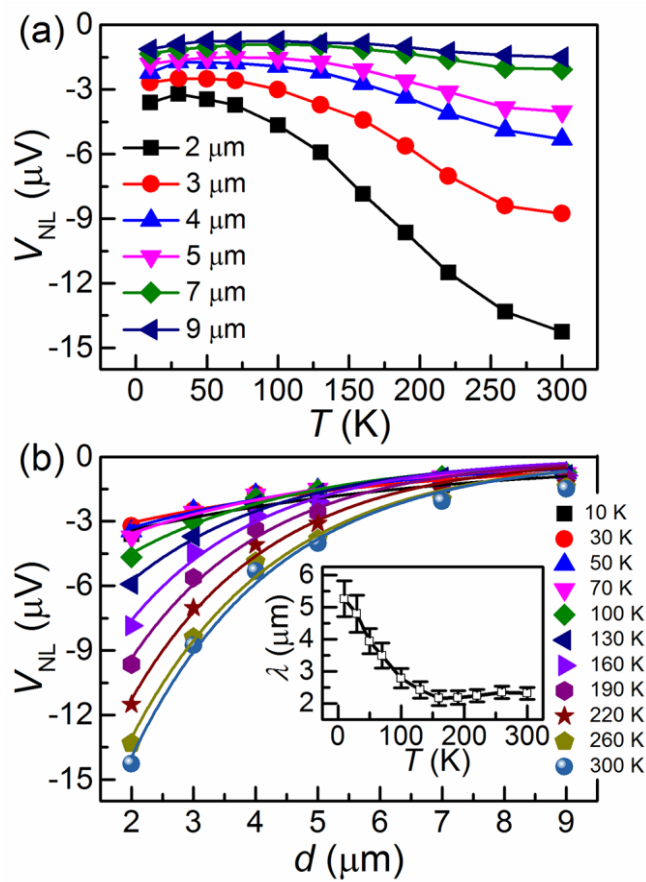

FIG. 4. Data from devices with 80 nm YIG thicknesses. (a) $V_{NL}$ versus $T$ at different separation distances $d$. (b) The same data reorganized as $V_{NL}$ versus $d$ at different temperatures. Curves in (b) are fit results to a single exponential decay. The inset in (b) is the magnon relaxation length extracted from the fits as a function of temperature.

Evaluating the systematically accumulated data, a physical model is needed to explain in particular the polarity reversal in $V_{NL}$ seen in thick film devices. It is very suggestive of competing transport processes delivering opposed spin magnons. The magnon chemical potential $\mu$ is introduced to evaluate the magnon spin accumulation and distribution through LSSE in the vicinity of the injector, depicted in Fig. 5(a). Magnons of opposite chemical potential ($\mu^+$ and $\mu^-$) are driven towards hotter and colder directions under the thermal gradient.[21] Opposed magnon spins accumulate at the YIG/Pt and YIG/GGG interfaces and the orientation flips direction in the bulk of the YIG film.[21,31,32]

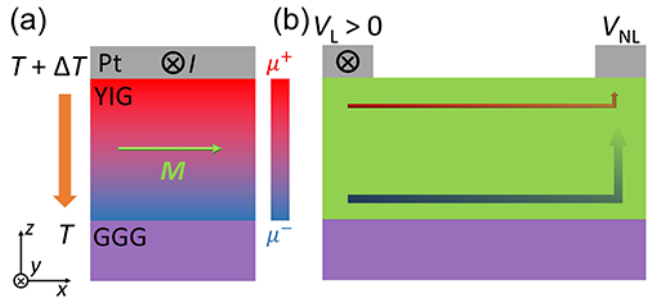

Fig. 5. (a) Diagram of magnon chemical potential $\mu$ distribution near the injector. (b) Sketch of the competing two-channel model for TEM transport in YIG. The top channel at the YIG surface and the bottom channel at the YIG/GGG interface are marked by red and blue arrows.

Inspired by the distribution of opposed thermal magnons and the surface mode of magnetostatic spin waves observed in in-plane-magnetized magnets,[33] a two-channel transport model is proposed. Although magnon transport occurs throughout the film,[13,19,20] the YIG top surface and the YIG/GGG interface are higher conductance channels as illustrated in Fig. 5(b), in which opposed-spin TEM transports to the detector independently. The net magnon concentration in the detector, measured as $V_{NL}$, is determined by the competition between the two channels. When the flux from the top surface channel exceeds that of the bottom interface channel, magnon accumulation in the detector is similar to the injector and $V_{NL}$ is positive like local thermal voltage. Conversely, negative $V_{NL}$ is detected when the bottom interface channel dominates.

An extra path length equivalent to the YIG thickness is added for TEM transport by the bottom interface channel. TEM transport by the top surface channel seems easier to be detected and positive $V_{NL}$ should be more extensive in thicker YIG films, consistent with the data in Fig. 2. However, as a corollary to this model, the role of interface quality is also critical.[34] Higher magnon transport efficiency may be expected in the bottom spin channel, benefitting from the excellent lattice matching at the YIG/GGG interface.[27] Furthermore, an estimation of magnon relaxation length of 2.13 μm in the top channel is smaller than 2.32 μm in the bottom channel at 300 K, verifying the higher transport efficiency in the bottom spin channel as the larger characteristic length and it provides a path to scale transport efficiency quantitatively. Hence, at low temperature the negative chemical potential at the bottom interface channel gains enough magnon flux to compensate for losses in the additional path, and overcomes its top surface channel opponent. This results in negative $V_{NL}$, which reverses sign as the temperature increases. Similarly, at larger $d/t$ ratios, the relative amount of extra path compared to the bottom channel decreases, yielding negative $V_{NL}$ at most or all temperatures. So far, the proposed two-channel TEM transport model qualitatively agrees with the observed dependence of spin information on temperature, YIG thickness, and separation distance. More quantitative calculations, such as characteristic length scales and transport efficiencies of the two modes, are worthy future areas of inquiry. Very recently an alternative explanation has been proposed by Shan *et al.*,[20] where a similar sign reversal is observed, qualitatively explained by bulk-driven SSE together with the magnon diffusion model. Considering the complex processes of TEM transport, further theoretical research is highly desirable.

In summary, we have experimentally observed a sign reversal in the lateral transport of thermally excited magnons in YIG. A comprehensive set of data has been taken on the temperature, YIG thickness, and separation distance dependence of the magnon transport sign in a non-local geometry. A proposed model posits competitive two-channel, opposed-sign TEM transport. The unique transport properties of thermal magnons in magnetic insulators is beneficial to the controllable magnon transport and would accelerate the development of magnon-based applications in the future.

**SUPPLEMENTARY MATERIAL**

See supplementary material for the thermal origin of voltage in the detector and injector.

**ACKONWLEDGEMENT**

We are grateful to Dr. J. Xiao for fruitful discussions. This work was supported by Ministry of Science and Technology of the People's Republic of China (Grant No. 2016YFA0203800) and the National Natural Science Foundation of China (Grant Nos. 51671110, 51571128, and 51231004).

**Supporting information**

**I. Thermal origin of the non-local voltage in the detector**

Measurements were conducted in the device with YIG thickness of 2.95 μm and separation distance of 2 μm at 10 K. As shown in Fig. S1(a), angular dependence of $V_{nl}$ is identical once YIG is saturated under different magnetic fields (500 Oe, 2 kOe, and 5 kOe). This steady angular relationship as verified in the 2 kOe situation is in accordance with the characteristic transport of thermally excited magnons in non-local geometry [S1]. Meanwhile, both the linear correlation between $V_{NL}$ and $I^2$ at discrete current and quadratic relation between $V_{nl}$ and $I$ with sweeping current, as displayed in Fig. S1(b), exhibit the typical feature of heating effect as $\nabla T \propto I^2$. So far, transport of thermally excited magnons is recognized as the main contribution to the non-local voltage in our experiments.

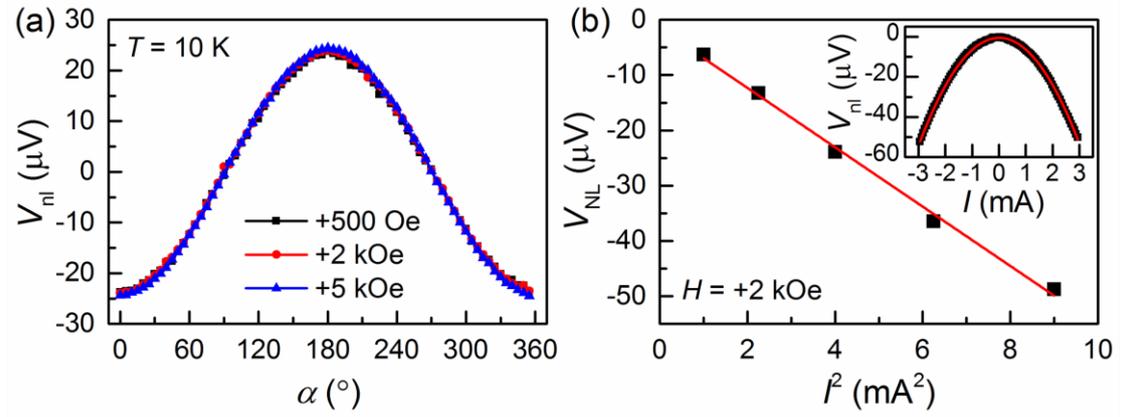

**FIG. S1.** (a) Angular dependence of the non-local voltage $V_{nl}$ measured with different in-plane magnetic fields applied. (b) Non-local saturated voltage $V_{NL}$ as a function of square of injected current $I^2$. The data is extracted from field-dependent measurements with different current and well fitted by a linear fitting (red line). Inset of (b) is the current dependence of $V_{nl}$ at a positive field of 2 kOe and a quadratic curve (red line)

is used to fit the data.

## II. Sign of local thermal voltage in the injector

As the source of thermally excited magnons, a local thermal voltage $V_l$ caused by spin Seebeck effect would generate concomitantly in the injector and this spin Seebeck voltage is obtained by current heating configuration [S2] in the local measurements. Considering the sign reversal of $V_{NL}$ is observed at $T_R$ (~75 K) [Fig. 1(d)], specific temperatures of 50 K and 100 K are selected at the two sides and a clear sign reversal of $V_{NL}$ is confirmed as revealed in Fig. S2(a). However, the situation is quite different in the local thermal voltage. As shown in Fig. S2(b), the magnitude of local thermal voltage $V_L$, defined as the saturated value at positive magnetic field, remains positive in both 50 K and 100 K. This unchanged sign of $V_L$ at different temperature is consistent with the previous work in GGG/YIG/Pt [S3] and we can determine the positive sign of local thermal voltage ($V_L > 0$) in the injector in our experimental geometry.

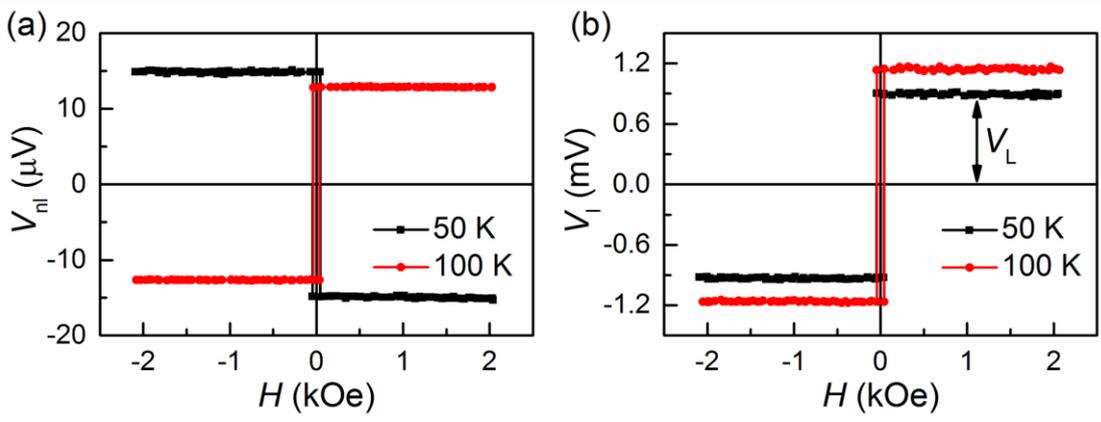

**FIG. S2.** (a) Field dependence of non-local voltage $V_{nl}$ in the detector. (b) Field dependence of local thermal voltage $V_l$ in the injector. The magnitude of local thermal voltage $V_l$ is defined as $V_L$, similar to the definition of $V_{NL}$. Representative

temperatures of 50 K and 100 K are selected at the two sides of $T_R$ (~75 K).